# Effect of Thermal and Mechanical Rejuvenation on Rheological Behavior of Chocolate

Tulika Bhattacharyya and Yogesh M Joshi*

Department of Chemical Engineering, Indian Institute of Technology Kanpur,

Kanpur, Uttar Pradesh, 208016

INDIA.

* Corresponding Author, email: joshi@iitk.ac.in

**Abstract**

Chocolate is known to undergo solid-liquid transition upon an increase in temperature as well as under application of deformation field. Upon sudden reduction in temperature from a molten state (or thermal rejuvenation), rheological properties of chocolate undergo evolution as a function of time under isothermal conditions, a behavior reminiscent of physical aging in polymeric glasses. Then again, subsequent to cessation of shear flow (or mechanical rejuvenation), chocolate shows temporal evolution of the rheological properties, a behavior similar to physical aging in soft glassy materials. In this work, we evaluate three rheological properties, namely, dynamic moduli, relaxation time spectrum and characteristic relaxation time of chocolate, and compare their evolution after thermal as well as mechanical rejuvenation. We observe that the evolution of the rheological properties subsequent to mechanical rejuvenation is distinctly different from that of thermal rejuvenation, wherein the evolution is more gradual in the former case. On the one hand, this work provides unique insights into how shear affects the rheological behavior of chocolate. On the other hand, this work clearly suggests that chocolate explores different sections of the energy landscape after mechanical rejuvenation compared to that of thermal rejuvenation.

**Introduction:**

Chocolate is a popular food product with remarkable sensory powers. Hot molten chocolate is a dense colloidal suspension with 65 – 75 mass % [1] of solid particles in a continuous matrix of fat stabilized by edible amphiphilic emulsifiers [2-6]. From a rheological point of view, the volume fraction and size distribution of the particulate matter along with the interparticle interactions among the same strongly influence the apparent viscosity of chocolate in a molten state [1, 4, 7-13]. With a decrease in temperature, part of the fat component of chocolate undergoes crystallization in different crystalline polymorphs forming several polycrystalline domains [2, 14]. Overall, a decrease in temperature, also known as thermal quench, enhances the viscosity of the chocolate, a behavior similar to that observed in amorphous or semicrystalline polymeric materials [15]. The solid particles suspended in the continuous phase may form space spanning percolated network or simply get jammed owing to high solid content [12]. Application of deformation/ flow field either breaks the structure or induces unjamming, causing yielding of the same [1, 4, 13, 15-19]. Application of deformation field, therefore, causes alteration of structure, a property that has been termed as rejuvenation, a behavior commonly reported for polymeric glasses [20-22], colloidal glasses and gels [23-27]. How chocolate is processed under the application of changing temperature and deformation fields strongly affects the structural and sensory properties of the final product [2, 28]. Therefore, it is vital to understand the effect of both these parameters. In this work, we investigate how the rheological properties, particularly the relaxation dynamics, evolve during structure formation in chocolate subsequent to thermal rejuvenation and compare the same with what happens after the mechanical rejuvenation.

Molten chocolate is a concentrated suspension of cocoa particles, sugar and milk powder in a continuous matrix of cocoa butter. The latter is primarily fat, which crystallizes upon cooling. Three major types of triglycerides are found in cocoa butter, namely, SOS, POP and SOP, where S, P and O denotes Stearic acid, Palmitic acid and Oleic acid, respectively [28-30]. These triglycerides can crystallize in six different polymorphs, named with roman numbers in their increasing order of melting point, Form I to Form VI. Particularly, oleic acid is unsaturated (has double bonds) and hence it is harder to crystallize. The melting point and type of chain packing for each of the polymorphs is available elsewhere [29, 31, 32]. Owing to the most stable form of chain packing, Form V and Form VI have the highest melting temperatures. Upon cooling of the chocolate from a molten state (temperature around 50 - 60 °C), the fat component starts to crystallize. For a given combination of Stearic acid, Palmitic acid and Oleic acid in cocoa butter, the type of crystal structure formation depends on the cooling rate and the final temperature. Slower cooling provides more time and hence leads to a more stable crystal state[33].

On rapid cooling of molten chocolate, fat components of cocoa butter crystallize to Form I. Owing to the unstable nature of Form I crystals, it rapidly transforms to Form II, which slowly transforms to Form III and Form IV. Inside the cooling tunnel of a chocolate manufacturing plant, untempered chocolate suspension (hot molten chocolate without any prior crystal) mostly eventually leads to the formation of Form IV crystals. However, if sufficient time is spent in the cooling tunnel, formation of Form V crystals has also been observed [29]. Form V type of crystal is the most desirable crystal state for the manufacturer due to its smooth texture, better heat stability and excellent surface finish. Form VI type of crystal gets developed after prolonged storage time of the order of weeks, however owing to its longer melting time inside the human mouth, its formation is undesirable for a manufacturer [2, 34]. To rapidly produce form V type of crystal in industries, a tempering process is performed on chocolate, wherein it is subjected to a varying temperature field [29, 31, 35].

Over the last few decades, the effect of shearing to modulate the structure has received much interest. The formation of shear induced phases in cocoa butter has been reported in literature [36-38]. Recent studies have found that shearing can accelerate the kinetics of crystallization. Shearing the chocolate mass breaks the crystallites and leads to the formation of higher number of seed crystals. Presence of a larger number of seed crystals allows more uniform crystal growth. Furthermore, triglyceride chains align themselves in the direction of shear and flow past each other, further enhancing the crystal growth. Stapley and coworkers [33] investigated the combined effect of shear rate and tempering time. Their results show that there is a critical shear rate required for seed crystal formation for any specified tempering time. Studies have shown that the time required for the onset of structure formation, i.e., induction time, is higher for the lower shear rate used during tempering [39]. They attribute this behavior to the large number of small size seed crystals formed at a higher shear rate. However, the rate of structure growth is reported to be slower for a higher applied shear rate. Dhonsi and Stapley [40] further show that the dependence of induction time on shear rate minimizes with the decrease in temperature.

In addition to crystalline moieties associated with cocoa butter, particulate matter comprising of cocoa particles, sugar, and milk powder is present in chocolate in a significant amount (65-75 mass %) [1]. In colloidal systems, such a high concentration of particulate matter has been observed to cause soft glassy dynamics [41]. Particularly, owing to high concentration of the particles, individual particles are arrested in cages formed by the surrounding particles restricting the mobility of the same significantly [41, 42]. It has also been reported in the literature that the particles may form a space spanning percolated network, with cocoa butter being the continuous phase [12]. Crystalline moieties of the cocoa butter, if present, may also become part of the network along with particulate matter [28]. Application of deformation field breaks the physical cages as well as the network (if any), inducing fluidity in the chocolate [1, 17-19], a behavior typically observed in thixotropic materials [38, 43, 44]. Such reduction in viscosity of the chocolate with time under application of deformation field (under isothermal conditions) is seemingly comparable to the reduction in viscosity with time caused by an increase in temperature. A qualitatively similar type of behavior has been reported for amorphous polymeric materials (polymer glasses) [15, 21, 45]. Overall, the chocolate shows remarkable similarity with soft glassy materials as well as polymeric glasses. Interestingly both types of glasses are known to show physical aging wherein constituents of the same undergo structural reorganization, wherein their relaxation time enhances as a function of time [21, 46]. Conversely, the application of deformation field decreases the relaxation time of both soft glasses as well as polymeric glasses, a behavior known as mechanical rejuvenation [47, 48]. In polymeric glasses increase in temperature beyond the glass transition temperature also causes a decrease in relaxation time, an effect known as thermal rejuvenation [21]. However, for polymeric glasses, it has been observed that mechanical rejuvenation may not have the same effect as thermal rejuvenation as both take the system to different sections of the energy landscape [47, 49]. The actual effect of mechanical rejuvenation could be so complex that for intermediate strengths of the deformation field, it could increase the relaxation time than decrease [50, 51].

It is, thus, apparent that chocolate shows peculiar similarity with both soft glassy materials as well as polymeric glasses that show physical aging and rejuvenation.

Furthermore, both temperature, as well as deformation fields are needed to impart desirable properties in chocolate. The objective of this work, therefore, is to study the effect of deformation field and temperature field on the rheological properties, particularly the relaxation dynamics of commercial chocolate. We also compare physical aging behavior after thermal as well as mechanical quench in the same.

**Materials and Methods:**

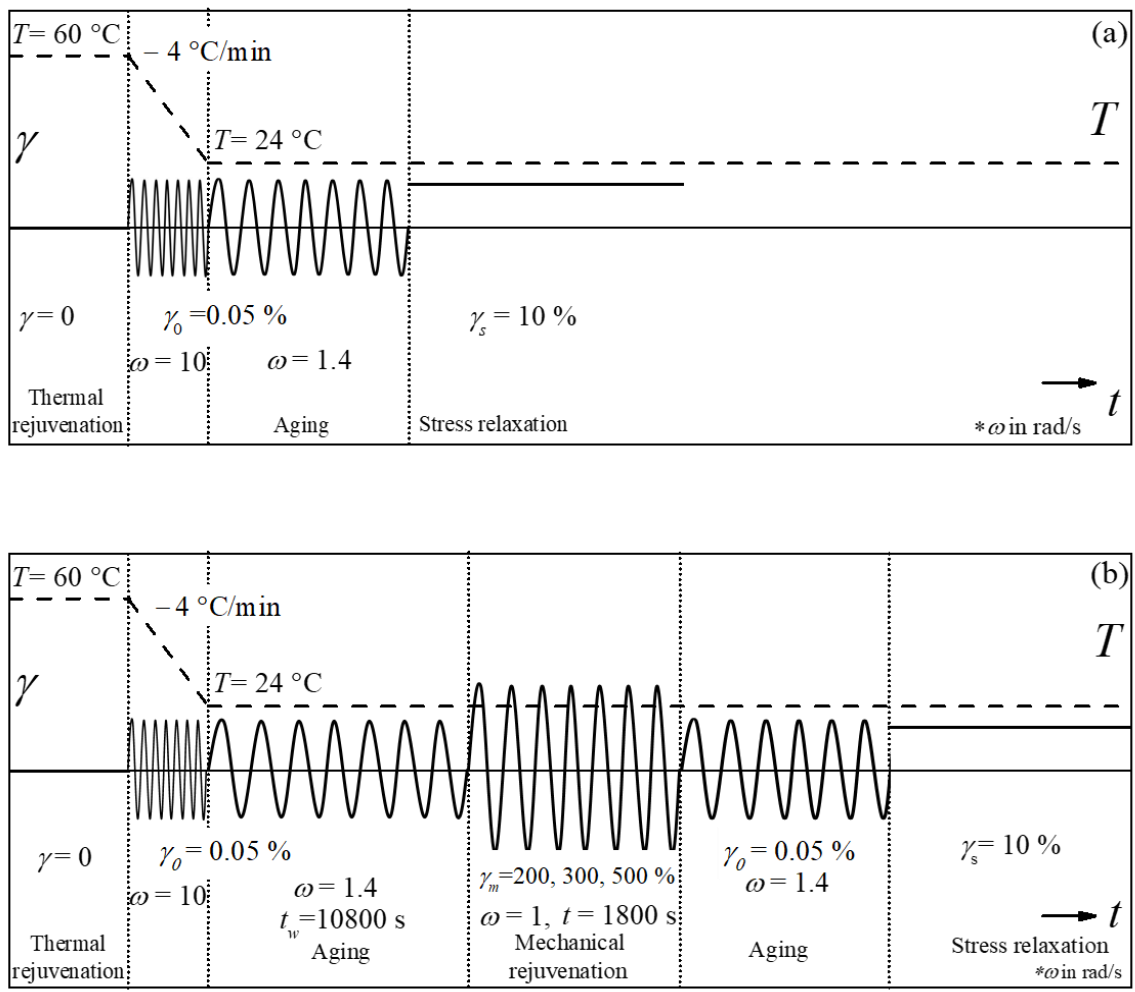

**Fig. 1** Experimental protocol for stress relaxation experiment on the chocolate sample at different time elapsed after thermal rejuvenation (a) and mechanical rejuvenation with different strain values (b).

In this work, we use commercially available Dairy Milk Silk chocolate manufactured by Cadbury. The procured bar contains sugar, approximately 25 % milk solids, cocoa butter, cocoa solids and emulsifiers 442 and 476. Emulsifier 442 either refers to soy lecithin or ammonium phosphatides. Emulsifier 476 suggests a mixture of glycerol and fatty acids and is popularly known as PGPR. The nutritional information provided by the manufacturer suggests that the specific chocolate contains 55.5 mass % sugar and 31.1 % fat. The chocolate bars are preserved in a refrigerator at around 4 °C temperature prior to use. Before rheological measurement, a chocolate bar is melted at 60 °C and simultaneously stirred gently using a spatula to facilitate homogenization during melting. The homogenized sample is subsequently loaded in the shear cell of a rheometer. In this work, we use a stress-controlled rheometer, Anton Paar MCR 501 (serrated concentric cylinder geometry with gap thickness 1.13 mm and outer cylinder radius 28.915 mm) for the rheological measurements. We use a uniform cooling protocol for all the experiments, wherein chocolate melt at 60°C is cooled to 24°C by applying a ramp of −4°C/min. In this work, we do two types of rheological experiments, which are performed after thermal as well as mechanical rejuvenation. In the first type of experiment, we subject material to small amplitude oscillatory shear (strain amplitude $\gamma_0$ of 0.05 % and angular frequency $\omega$ of 1.4 rad/s.) at 24°C subsequent to either of the rejuvenation and monitor the evolution of elastic and viscous moduli ($G'$ and $G''$) as a function of time. In the second type of experiment, we apply step strain to chocolate at different times elapsed since the cessation of either type of rejuvenation and assess the resultant relaxation dynamics. The typical rheological experimental protocol for the second type of experiment is shown in Fig. 1. As shown in Fig. 1(a), after the thermal quench to 24°C, the chocolate sample is subjected to step strain ($\gamma_s = 10$ %) at different times ($t_w$, aging times), and the corresponding relaxation of stress is monitored. The protocol shown in Fig 1(b) pertains to studying relaxation dynamics after mechanical rejuvenation. In this protocol, a sample is aged for 10800 s and subsequently, oscillatory strain amplitudes of $\gamma_m = 200, 300$ and 500 % having angular frequency $\omega$=1 rad/s is applied for 1800 s. Followed by cessation of mechanical rejuvenation, step strain experiments ($\gamma_s = 10$ %) are carried out at different $t_w$, and ensuing relaxation of stress is monitored.

In this work, we monitor evolution of viscoelastic behavior of chocolate at single frequency. However, in the literature many techniques have been proposed that can be categorized as Fourier rheometry or Fourier transform rheology, wherein rather than applying an oscillatory input at single frequency, an input with superposition of several frequencies is applied and the response is suitably resolved to obtain dynamic moduli[52-58]. The benefit of such techniques is ability to obtain a response over broad frequency range within an optimum time. Recently McKinley and coworkers[59, 60] optimized this concept and proposed a Chirp protocol that maximizes the signal-to-noise ratio by minimizing spectral leakage for fast time-evolving systems. The characterization of the present system may benefit from application of such Fourier transform rheological techniques to obtain information over a broad spectrum of frequencies.

**Results and Discussion:**

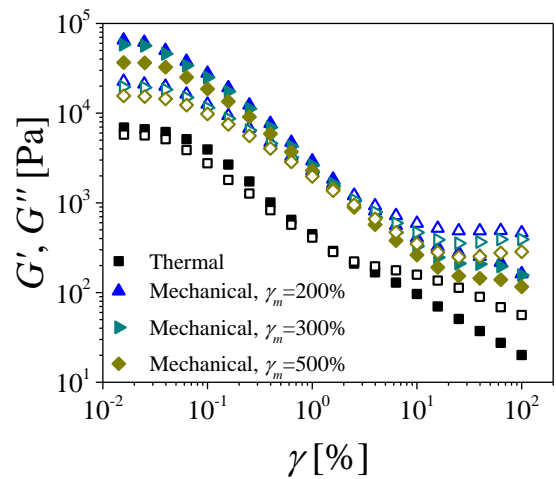

**Fig. 2** Evolution of $G'$ (closed symbols) and $G''$ (open symbols) plotted as a function of strain amplitude at $\omega$=1.4 rad/s after the thermal rejuvenation (squares) and mechanical rejuvenation with different strain amplitudes (up triangles: $\gamma_m$=200 %, right triangles $\gamma_m$=300 %, diamonds $\gamma_m$=500 %).

The chocolate sample at 60°C is in the molten state. We cool it in the rheometer to 24°C by applying a decreasing temperature ramp of 4°C/min. Soon after the sample is thermally equilibrated at 24°C, we subject it to oscillatory strain sweep with angular frequency, $\omega$ = 1.4 rad/s. The corresponding dependence of $G'$ and $G''$ on the amplitude of strain is plotted in Fig. 2. It can be seen that both the moduli are almost constant in the linear viscoelastic domain with $G' > G''$. With an increase in strain amplitude $G'$ decreases at a faster rate and eventually $G''$ dominates as usually observed in the strain sweep experiments. Fig. 2 also leads to the linear viscoelastic domain associated with strain. In an independent experiment, after thermal quench to 24°C, we subject the chocolate sample to small amplitude oscillatory shear with a strain magnitude of $\gamma_0 = 0.05$ % and angular frequency $\omega = 1.4$ rad/s. Fig. 3 shows the time evolution of $G'$ and $G''$ at 24°C. It can be seen that $G'$ and $G''$ evolve with aging time in four stages. In the initial stage, up to around 1000 s, both the moduli show a slow increase with the aging time. In this regime $G'$ increases more rapidly than $G''$ (stage 1). Subsequently, $G'$ and $G''$ undergo a sharp increase for next 2000 s (stage 2) with $G'$ always remaining greater than $G''$. For times from $t_w \approx 3000$ s to 11000 s, $G'$ and $G''$ show a near plateau behavior with $G''$ greater than $G''$ by around a factor of 4 (stage 3). Beyond $t_w > 11000$ s, yet again, $G'$ and $G''$ undergo a clear increase with time (stage 4). It should be noted that the strain of 0.05 % is at the threshold of linear and non-linear regime in the beginning of the aging process. However, soon material ages and the dynamic moduli of the same increase so significantly that this strain limit falls in the linear domain. Usage of strain lower than 0.05 % causes data to show more fluctuations at high aging times, where the material becomes stiff due to aging.

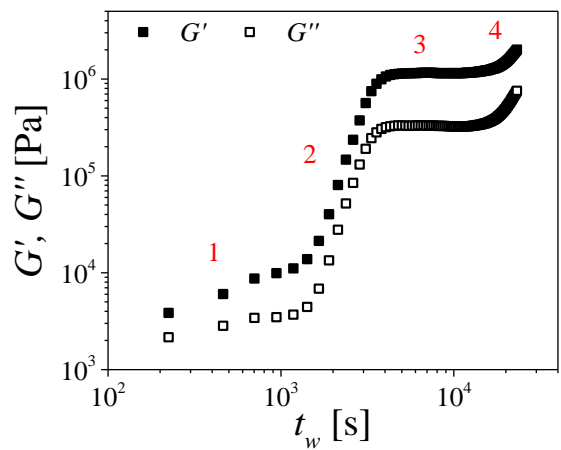

**Fig. 3** Evolution of $G'$ (closed symbols) and $G''$ (open symbols) plotted as a function of time at for $\gamma_0 = 0.05$ %, $\omega = 1.4$ rad/s and $T = 24°C$ after thermal quench.

Next, we explore the effect of application of mechanical rejuvenation, wherein at $t_w$=10800 s after thermal rejuvenation, we subject the chocolate to oscillatory strain of $\gamma_m$= 200, 300 and 500 % at angular frequency $\omega$ = 1 rad/s for 1800 s. In Fig. 4, we show how $G'$ and $G''$ evolve with time under application of mechanical rejuvenation. Before the sample is subjected to mechanical rejuvenation, as shown in Fig 3, $G'$ is greater than $G''$. After application of mechanical rejuvenation $G'$ seems to be rapidly decreasing below the $G''$ within the first few cycles (we apply the mentioned magnitude of strain at the angular frequency of $\omega$= 1 rad/s while the first point is recorded at 60 s). As expected, a decrease in both the moduli is more severe for a higher magnitude of strain. Subsequently, both the moduli continue to decrease in almost parallel fashion on the semilogarithmic scale, with curves associated with a higher magnitude of mechanical rejuvenation shifting downward. In a limit of long times, $G'$ and $G''$ attain a plateau as shown in Fig. 4.

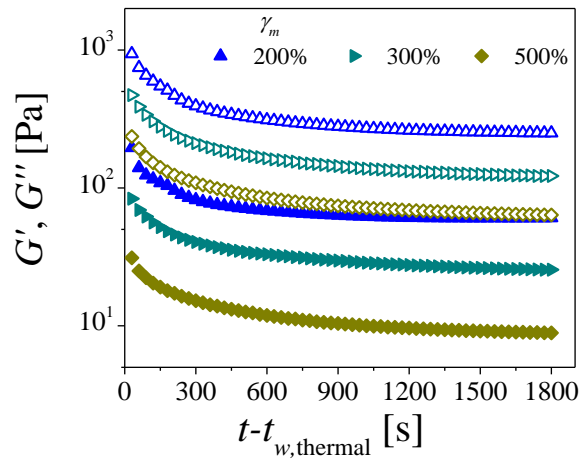

**Fig. 4** Evolution of $G'$ (closed symbols) and $G''$ (open symbols) as a function of time at $\omega$=1 rad/s during mechanical rejuvenation after physical aging at $T = 24°C$ for 10800 s. The different symbols represent mentioned values of $\gamma_m$ used for mechanical rejuvenation (up triangles: $\gamma_m$=200 %, right triangles $\gamma_m$=300 %, diamonds $\gamma_m$=500 %).

Immediately after the mechanical rejuvenation, we perform strain sweep on the chocolate samples at $\omega$=1.4 rad/s angular frequency. The corresponding dependence of dynamic moduli on strain is plotted in Fig. 2. In a limit of small strain, the dynamic moduli associated with all the three mechanical rejuvenation protocols show a constant value suggestive of linear viscoelastic limit. In this limit, we observe $G' > G''$. However, with increase in strain, both the moduli decrease with $G'$ decreasing more strongly than $G''$ such that the latter becomes larger than the former. Irrespective of the magnitude of strain, dynamic moduli associated with lower rejuvenation strain shift to a higher value. Overall, the values of dynamic moduli after mechanical rejuvenation are higher than those associated with that after thermal rejuvenation. At further high strain $\gamma_0 > 10\ \%$, both the moduli again show a plateau before eventually decreasing (such plateau is also observed for $G''$ for a thermally rejuvenated sample at around the same value of strain as shown in Fig. 2). This trend is a characteristic feature of two step yielding behavior reported for a wide range of soft materials [61-64]. Such behavior is attributed to the presence of structure over two broadly separated length-scales or when the first yielding event leads to certain structure formation that breaks down during the second yielding process [64, 65]. However, the investigation of two-step yielding in chocolate is beyond the scope of the present work.

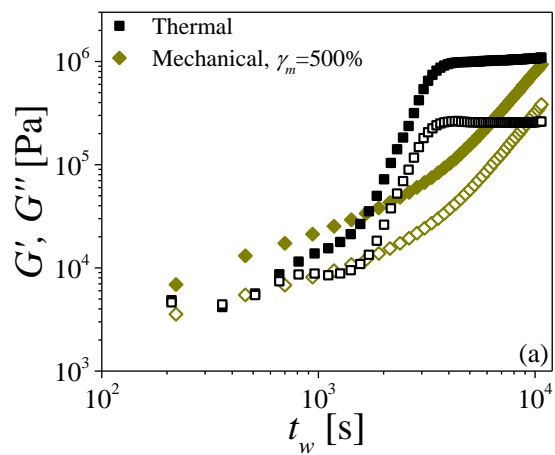

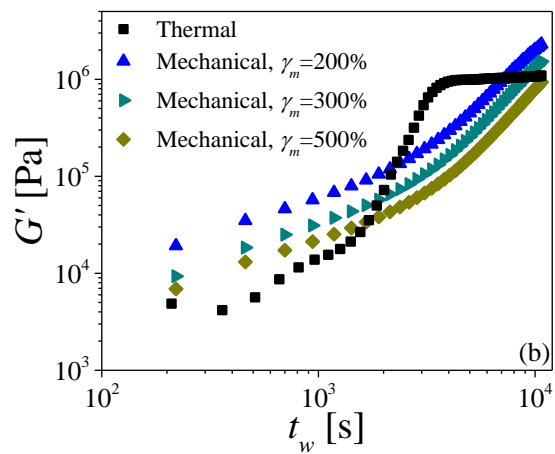

**Fig. 5** (a) Evolution of $G'$ (closed symbols) and $G''$ (open symbols) as a function of time at $\omega$=1.4 rad/s and at $T = 24$°C after thermal quench (squares) and mechanical quench at $\gamma_m$=500 % (rhombuses). (b) Evolution of $G'$ (closed symbols) as a function of time at $\omega$=1.4 rad/s and at $T = 24$°C after thermal quench (squares) and mechanical quench with different $\gamma_m$ (up triangles: $\gamma_m$=200 %, right triangles $\gamma_m$=300 %, diamonds $\gamma_m$=500 %).

In independent experiments, we explore the evolution of the dynamic moduli subsequent to mechanical rejuvenation. In Fig. 5a, we plot $G'$ and $G''$ as a function of time for a sample, for which rejuvenation was performed using strain magnitude of $\gamma_m$= 500 %. As a guide, we also plot the evolution of dynamic moduli after thermal rejuvenation on the same figure. It can be seen that the evolution of dynamic moduli subsequent to mechanical rejuvenation is more gradual compared to that after the thermal rejuvenation. During the entire evolution, $G'$ always remains at a higher level than $G''$ in such a way that $G'$ and $G''$ remain nearly parallel to each other. In Fig. 5b, we plot the evolution of $G'$ subsequent to mechanical rejuvenation at all the studied strains and compare the same with that obtained after thermal rejuvenation. We observe that on cessation of deformation field, $G'$ rapidly increases. At initial aging time, $G'$ of mechanically rejuvenated samples with $\gamma_m$= 200 and 300 % strain are always higher than the system rejuvenated with 500 % strain as well as that of a thermally rejuvenated sample. We observe the magnitude of $G'$ during physical aging is higher for lower magnitude of $\gamma_m$ and it remains so throughout the aging process with evolution of $G'$ of three mechanically rejuvenated systems remaining almost parallel to each other.

To investigate the evolution of relaxation dynamics during aging after the various rejuvenation protocols used, we subject the four types of samples to step strain ($\gamma_s$) at different $t_w$ since rejuvenation. Typical behavior of strain applied by rheometer and the corresponding evolution of stress is plotted in appendix for three aging times for a thermally rejuvenated system (Fig. A1) and mechanical rejuvenated system with $\gamma_m$=500% (Fig. A2). In Figs A1 and A2, it can be seen that owing to instrument inertia it takes around $\mathcal{O}(0.1 \text{ s})$ to reach the desired order of magnitude of strain. Interestingly, immediately after the step strain is applied, stress induced in a material first increases over a very short period and then starts decreasing. However, in a fast relaxation process, stress partly decays to an intermediate value over the same period required for strain to achieve the preassigned value. Such fast decay in stress is reminiscent of fast decay in glassy materials usually associated with the microscopic motion of the constituents within their cages [66, 67]. The relaxation subsequent to the achievement of the constant value by the strain is very slow. The first part of the relaxation is extremely fast, to be accurately monitored by rheometer. Furthermore, since strain has not reached its ultimate value and the relaxation dynamics seems to be qualitatively different over the fast relaxation process of $\mathcal{O}(0.1\text{ s})$ than the subsequent slow process [66, 67], we omit the analysis of the fast process. In figure 6 we plot relaxation modulus ($G = \sigma/\gamma_s$) for the thermally rejuvenated system only after the strain has attained the preassigned value, as a function of time elapsed since application of step strain ($t - t_w$) at different $t_w$ since application of thermal rejuvenation. We represent the value of relaxation modulus at which the desired value of strain is attained to be $G_s$. As expected, relaxation modulus decreases with time. We observe that for $t_w < 4000$ s, $G_s$ increases and the relaxation modulus curves shift to higher levels with increase in $t_w$. However, for $t_w > 4000$ s, $G_s$ is observed to be independent of $t_w$. This behavior matches well with the trend reported for $G'$ in Fig 5. Furthermore, the relaxation seems to be comparatively faster for $t_w < 3000$ s, as all the curves bend sharply downward at higher $t - t_w$. But at higher times the relaxation dynamic is significantly slower.

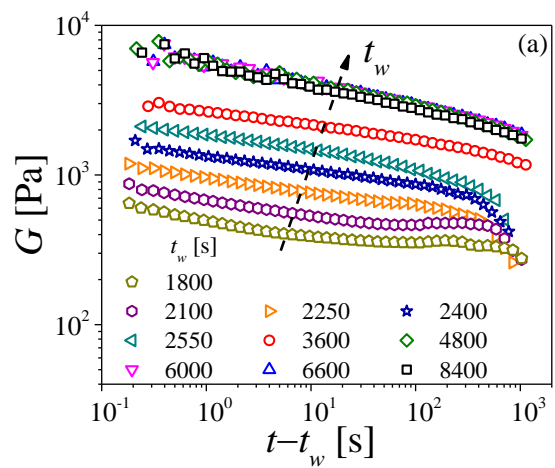

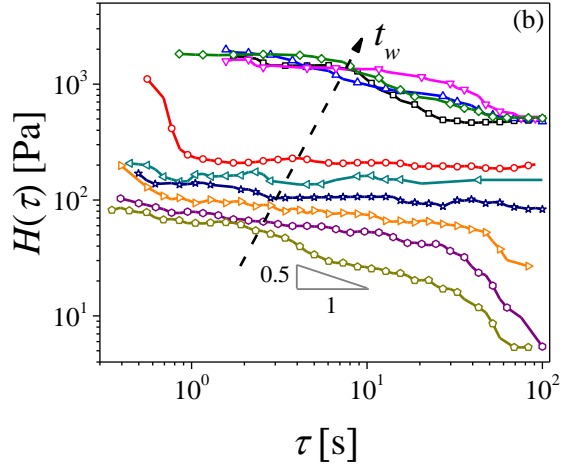

**Fig. 6** Stress relaxation modulus (a) and the corresponding relaxation time spectrum (b) is plotted as a function of time elapsed since application of step strain for a thermally rejuvenated sample. The different symbols denote the various $t_w$ after thermal rejuvenation. The arrow is a guide to the eye in the increasing direction of $t_w$.

Further insight into evolving relaxation dynamics can be obtained by plotting relaxation spectra $H(\tau)$ at different waiting times. The continuous relaxation spectrum $H(\tau)$ is defined in such a fashion that $H(\tau)d\ln\tau$ corresponds to infinitesimal contribution to modulus from relaxation time modes lying in between $\tau$ and $\tau + d\tau$ leading to relaxation modulus given by [68]:

$$G(t) = G_e + \int_{-\infty}^{\infty} H(\tau)\exp(-s/\tau)d\ln\tau, \qquad (1)$$

where $G_e$ is a discrete contribution to the spectrum at $\tau = \infty$, and $s$ is time. According to Ferry [69], Eq. (1) can be inverted to obtain an approximate relation:

$$H(\tau) \cong -\left(\frac{dG(s)}{d\ln s}\right)_{s=\tau}. \qquad (2)$$

We obtain $H(\tau)$ from the relaxation modulus data shown in Fig 6(a). However, since material is undergoing time evolution, we restrict to $s \equiv t - t_w$=100 s while computing $H(\tau)$ that gives instantaneous relaxation spectra over a limited relaxation modes. In Fig. 6(b), we plot $H(\tau)$ for the relaxation data shown in Fig 6(a). It can be seen that $H(\tau)$ has a negative slope ($\approx -0.5$ on a double logarithmic scale for the smallest explored $t_w$), which suggests that the faster the relaxation mode (smaller $\tau$) is, the greater is its dominance. Such behavior is reminiscent of a network structure [26, 70, 71]. With increase in $t_w$, the magnitude of slope of $H(\tau)$ versus $\tau$ decreases and the curves becomes progressively flat. This phenomenon suggests the increasing contribution of the slow modes with increase in $t_w$, which could be due to increase in network density. In a limit of large $t_w > 4000$ s, same as what observed for $G(t)$, $H(\tau)$ only shows a weak dependence on waiting time $(t - t_w)$ over the explored time duration. The fact that $H(\tau)$ remains practically unaffected for $t_w > 4000$ s, indicates extremely slow evolution of network density over the assessed period.

The corresponding relaxation behavior after mechanical rejuvenation for $\gamma_m$=500 % strain amplitude is plotted in figure 7(a). The corresponding relaxation behavior after $\gamma_m$ =200 and 300 % strain is plotted in appendix Figs. A3(a) and (c) respectively. In these figures, relaxation modulus has been plotted only after the strain has attained the preassigned value. For these systems, the data is shown for only selected values of $t_w$ to avoid cluttering. Contrary to thermal rejuvenation and irrespective of the magnitude of strain imposed during mechanical rejuvenation, we observe a gradual increase in $G_s$ for all the explored values of $t_w$. The corresponding relaxation modulus curves shift to higher levels with an increase in $t_w$. Unlike the relaxation modulus behavior after thermal rejuvenation that shows a clear downturn of $G(t)$, the behavior subsequent to mechanical rejuvenation shows a plateau with a slight upturn in a limit of large $t - t_w$. Plateau in the relaxation modulus is a characteristic feature of residual stresses in a system. On the other hand, a slight upturn in a relaxation modulus curve suggests an increase in modulus as a function of time, which could be due to physical aging in the present system [72]. Experimentally, such upturn in relaxation modulus has been reported in the literature [73]. In fig 7(b), we plot $H(\tau)$ for the stress relaxation described in Fig 7(a) for the system subjected to mechanical rejuvenation of $\gamma_m$= 500 % strain. It can be seen that $H(\tau)$ shows a power law dependence on $\tau$ with a negative slope, which is around $-0.7$ on a double logarithmic scale for the lowest explored $t_w$. With increase in $t_w$, magnitude of slope decreases similar to that observed subsequent to the thermal rejuvenation. However, the decrease in magnitude of slope subsequent to mechanical rejuvenation is more gradual and does not lead to saturation as observed for the thermally rejuvenated system over the explored time domain. As discussed before, the presence of a negative slope suggests network like structure and a decrease in its magnitude indicates an enhancement in the network density [26, 70, 71]. Clearly, this dynamics is more gradual subsequent to mechanical rejuvenation than that of after thermal rejuvenation. The corresponding relaxation time spectrum $H(\tau)$ after $\gamma_m$ = 200 and 300 % strain is respectively plotted in appendix figs. A3(b) and (d). Apart from the difference that both relaxation modulus and spectrum curves at respective times moves to higher values with a decrease in magnitude of $\gamma_m$, the overall behavior is qualitatively similar to that observed for a mechanically rejuvenated system with $\gamma_m$=500 % strain.

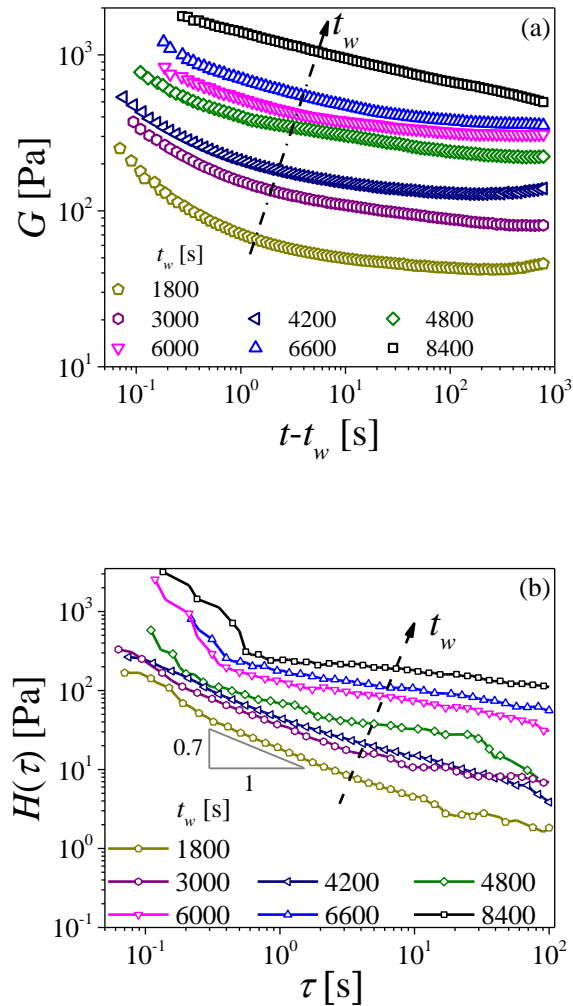

**Fig. 7** Stress relaxation modulus (a) and the corresponding relaxation time spectrum (b) is plotted as a function of time elapsed since application of step strain for a mechanical rejuvenated system with $\gamma_m$=500 %. The different symbols denote the mentioned waiting times $t_w$ after mechanical rejuvenation. The arrow is a guide to the eye in the increasing direction of $t_w$.

In Fig. 8(a), we plot relaxation modulus as a function of $t - t_w$ at $t_w$=6000 s for different rejuvenation protocols. It can be seen that value of $G_s$ is maximum for the thermal rejuvenation protocol. Furthermore, $G_s$ decreases with an increase in the applied magnitude of $\gamma_m$. The corresponding relaxation spectrum $H(\tau)$ is plotted in Fig 8(b). As expected, all the $H(\tau)$ curves depict a negative slope, but the magnitude is lower with a decrease in extent of mechanical rejuvenation. For the system rejuvenated with 200 % strain $H(\tau)$ is practically flat which is suggestive of denser network. The thermally rejuvenated system at $t_w$=6000 s, while shows negative slope, the overall magnitude of $H(\tau)$ is quite high for this system, suggestive of qualitatively different dynamics.

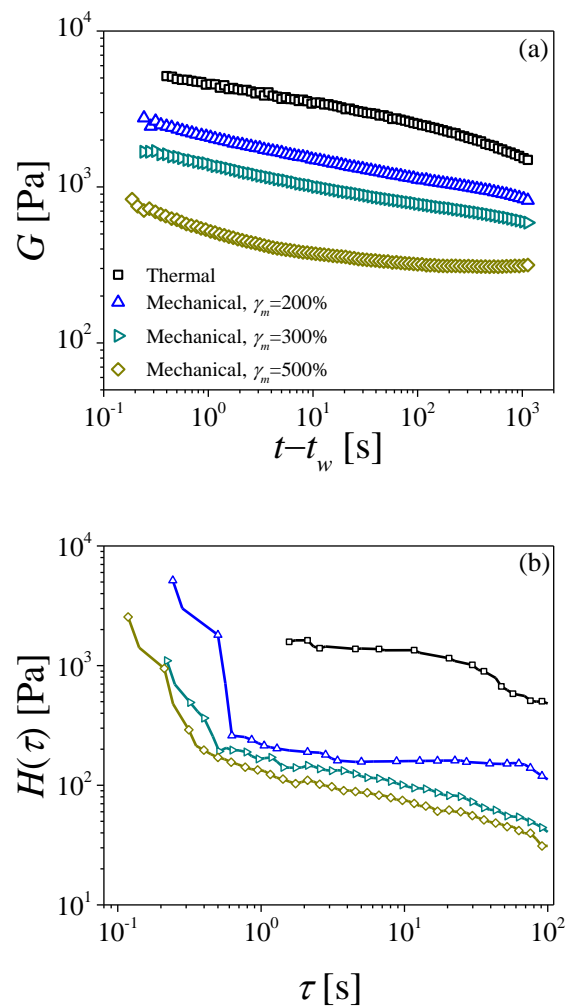

**Fig. 8** The stress relaxation modulus (a) and the corresponding relaxation time spectrum (b) is plotted as a function of time elapsed since application of step strain at $t_w$= 6000 s for the thermally rejuvenated sample (squares) as well as mechanical rejuvenated

samples with different $\gamma_m$ (up triangles: $\gamma_m$=200 %, right triangles $\gamma_m$=300 %, diamonds $\gamma_m$=500 %).

To characterize the stress relaxation behavior, we obtain the time required for $G_s$ to reach half of its value $(t_{1/2})$ following the time evolution curve of $G(t - t_w, t_w)$. We take $t_{1/2}$ as a measure of the characteristic relaxation time and plot $t_{1/2}$ as a function of $t_w$ in figure 9 for all the four explored rejuvenation protocols. In case of thermally rejuvenated system, $t_{1/2}$ shows a very steep increase with $t_w$ upto 3000 s and then shows a weak increase with an increase in $t_w$ for higher values. Such a two-step behavior in the evolution of characteristic relaxation time can also be observed for mechanically rejuvenated samples, but its nature is qualitatively different than thermally rejuvenated system. At any $t_w$, $t_{1/2}$ is highest for thermally rejuvenated system and decreases with an increase in the applied magnitude of $\gamma_m$ during mechanical rejuvenation. It can be seen that, the nature of evolution of $t_{1/2}$ for all the rejuvenation protocols, closely resembles the corresponding evolution of dynamic moduli shown in Fig. 5. Therefore, rheological behavior in general and relaxation dynamics in particular, very clearly suggests profound impact of the rejuvenation protocol on the studied chocolate system.

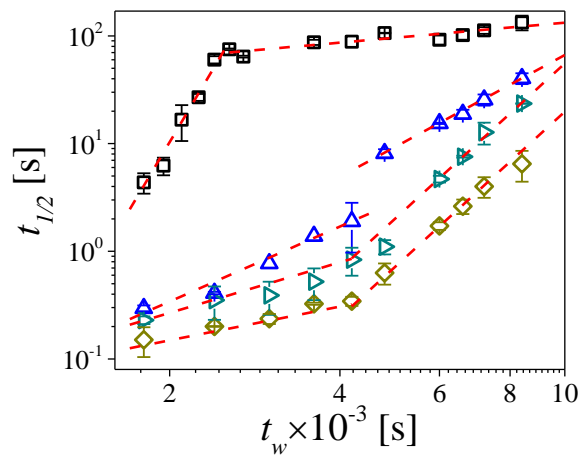

**Fig. 9** The characteristic relaxation time $t_{1/2}$ is plotted as a function of $t_w$ for the thermally rejuvenated system (squares) and the mechanically rejuvenated systems with different $\gamma_m$ (up triangles: $\gamma_m$=200 %, right triangles $\gamma_m$=300 %, diamonds $\gamma_m$=500 %). The dashed lines are a guide to the eye.

The rheological observation discussed in the present work can be correlated with microstructural changes occurring in chocolate. However, in this work we focus only on rheological behavior, and analogy to possible microstructure is conjectured only through the reference to literature. Upon cooling since thermal rejuvenation, randomly arranged fatty acid chains start nuclei formation, and for temperatures below ≈28°C chains pack closely with each other. This leads to crystalline domain formation in the continuous phase [29, 31, 33, 39]. The evolution of solid fat content (crystallized fat) from NMR spectroscopy under isothermal conditions over a similar temperature range has been widely reported to show a stepwise increase. Interestingly, the trend of time evolution curve of $G'$ and $G''$ shown in Fig 3, upto stage 3 qualitatively matches well with evolution of crystalline fat content observed in the literature [32, 74, 75]. Owing to this, the observed increment in dynamic moduli seem to be related to formation of crystalline moieties in the continuous phase of fatty acid chains. According to Marangoni et al. [32], over a temperature range of $21 - 26$ °C, cocoa butter crystallizes directly to Form IV under isothermal conditions, and the process of crystalline domain formation eventually saturates with aging time. It could be further possible that the particulate matter, along with the crystallizing domains of fat, forms a space spanning network soon after the thermal quench. Various studies also propose possibility of such percolated network formation [1, 12], which could explain the negative slope of the relaxation time spectrum at early times shown in Fig. 6b. The nature of network formation majorly depends on the packing ability of the fatty acid chains present in the suspension. As time passes, more material crystallizes, as suggested by Marangoni et al. [32], and the magnitude of the slope rapidly decreases to show a flat relaxation time spectrum.

Subsequent to aging chocolate at 24°C for three hours, we subject the same to oscillatory strain fields of varying magnitude. During this process, the crystallites break and get oriented in the direction of shear [33, 36]. Orientation of the crystalline network of the fatty acid chains in a particular direction during mechanical rejuvenation is otherwise absent after thermal rejuvenation. It is expected that with an increase in the intensity of strain applied during the mechanical rejuvenation stage, greater will be the breakage of large crystalline domains, and more will be the orientation survived domains [39]. This explains the decrease in the plateau value of dynamic moduli with an increase in magnitude of oscillatory strain applied during mechanical rejuvenation, as shown in Fig. 4. Owing to breakage of crystallites, after cessation of mechanical rejuvenation, chocolate is expected to have greater number of seed nuclei, which have been proposed to undergo more homogeneous crystallization leading to stable crystal forms that possess a refined grain size. MacMillan et al. [36] suggested that shearing breaks the van der Waals forces acting in lower forms of fat crystals and aids in better packing of fatty acid chains. This proposal suggests the gradual growth of dynamic moduli subsequent to mechanical rejuvenation. Since the greater intensity of oscillatory strain causes greater breakage of the original structure, the evolution of dynamic moduli shifts to lower values with an increase in magnitude of $\gamma_m$ as shown in Fig. 5b. The magnitude of negative slope of relaxation time spectra is also observed to be decreasing gradually with time for the mechanically rejuvenated systems. In addition, the relaxation time of the mechanically rejuvenated systems is significantly smaller than that subsequent to thermal quench suggests that not just the leaner network after mechanical rejuvenation but densification of the network to be also a slower process compared to what is observed subsequent to thermal rejuvenation.

Evolution of dynamic moduli and relaxation dynamics subsequent to thermal rejuvenation and mechanical rejuvenation are distinct in nature. Similar behavior has been proposed for polymeric glasses with respect to how rejuvenation occurs subsequent to thermal quench in comparison with mechanical quench [20, 47]. Similar to what is proposed for polymeric glasses, we also attribute the difference between the two rejuvenation protocols used to their uniqueness of how the energy landscape is explored. The difference in alteration of microstructure corresponding to each rejuvenation protocol forces the system to explore different sections of the phase space, leading to stark differences in evolution of rheological properties [76]. We believe that such behavior is not limited to chocolates but could be observed in food products such as Peanut Butter, Jaggery, cheese, etc., wherein solid-liquid transition takes place with an increase in the temperature as well as the intensity of deformation field.

**Summary**

Chocolate belongs to that class of materials, which undergo solid-liquid transition upon increase in temperature as well as under application of deformation field.

Subsequent to thermal as well as mechanical quench, under isothermal conditions, the rheological properties of chocolate undergo time dependent change. This behavior of chocolate is suggestive of physical aging observed in molecular glasses that undergo thermal rejuvenation as well as soft glassy materials that undergo mechanical rejuvenation. In the food industry, chocolate is usually subjected to both temperature as well as shear fields in a controlled fashion in order to get the desired properties. In this work, we study the time evolution of three rheological properties, namely, dynamic moduli, relaxation time spectra and characteristic relaxation time subsequent to thermal and mechanical rejuvenations. In general, we observe that the evolution of the rheological properties is more gradual subsequent to mechanical rejuvenation compared to that after the thermal rejuvenation. We also study the effect of intensity of mechanical rejuvenation and observe that application of greater amplitude of oscillatory strain shifts the evolution of dynamic moduli as well as characteristic relaxation time to the lower magnitudes. It is also apparent that the evolution of rheological properties in these systems follows starkly different path, with magnitude of relaxation time orders of magnitude smaller in case of mechanical rejuvenation compared to that after the thermal rejuvenation. This behavior suggests that the system explores very different sections of the phase space in both types of rejuvenation protocols.

**Appendix:**

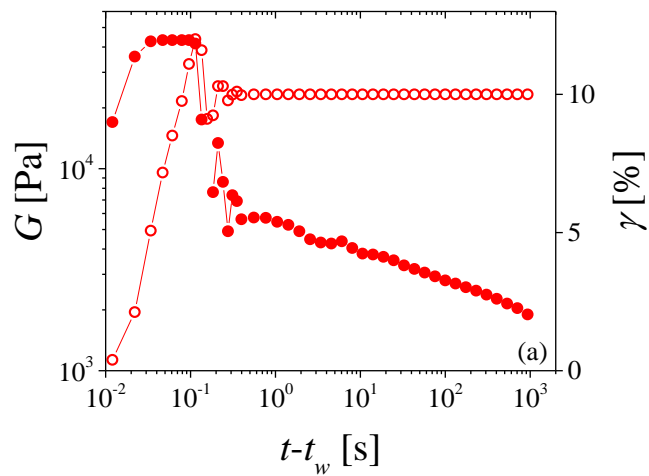

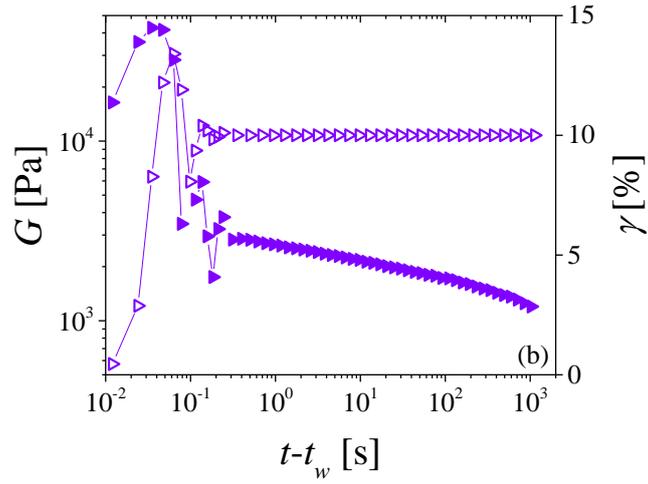

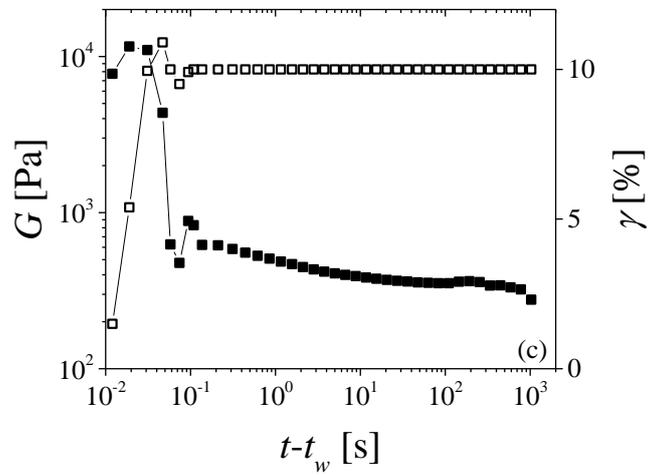

**Fig. A1** Stress relaxation modulus (closed symbols) and induced strain (open symbols) are plotted as a function of time elapsed since application of step strain for a thermally rejuvenated sample for various $t_w$ after thermal quench (a) $t_w$=7200 s, (b) $t_w$=3600 s, (c) $t_w$=1800 s.

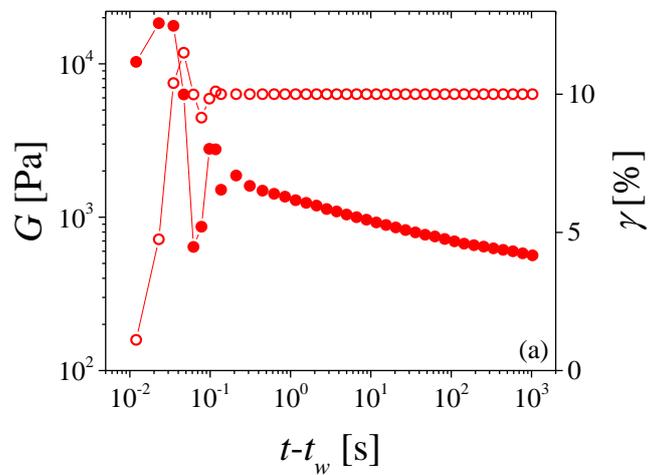

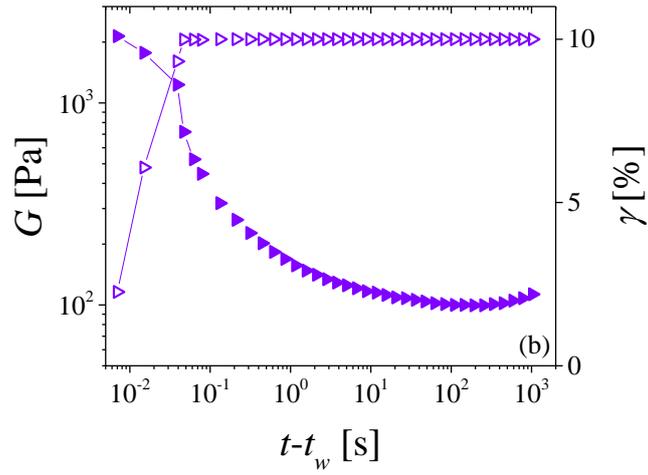

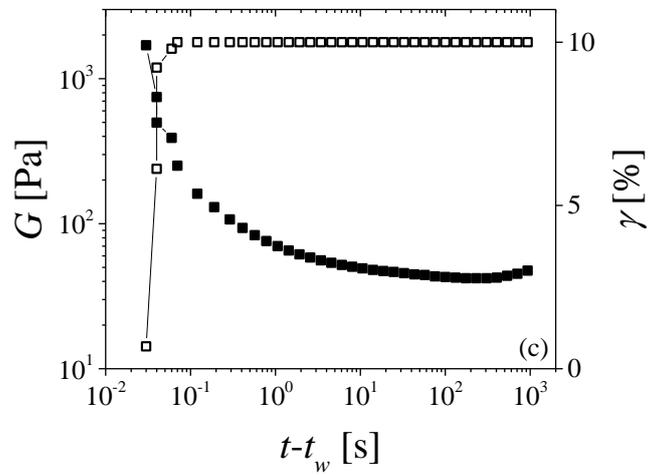

**Fig. A2** Stress relaxation modulus (closed symbols) and induced strain (open symbols) are plotted as a function of time elapsed since application of step strain for a mechanically rejuvenated sample with $\gamma_m$=500 % for various $t_w$ after mechanical quench. (a) $t_w$=7200 s, (b) $t_w$=3600 s, (c) $t_w$=1800 s.

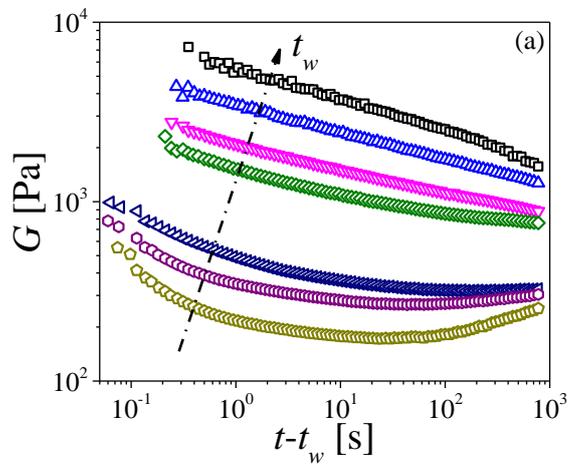

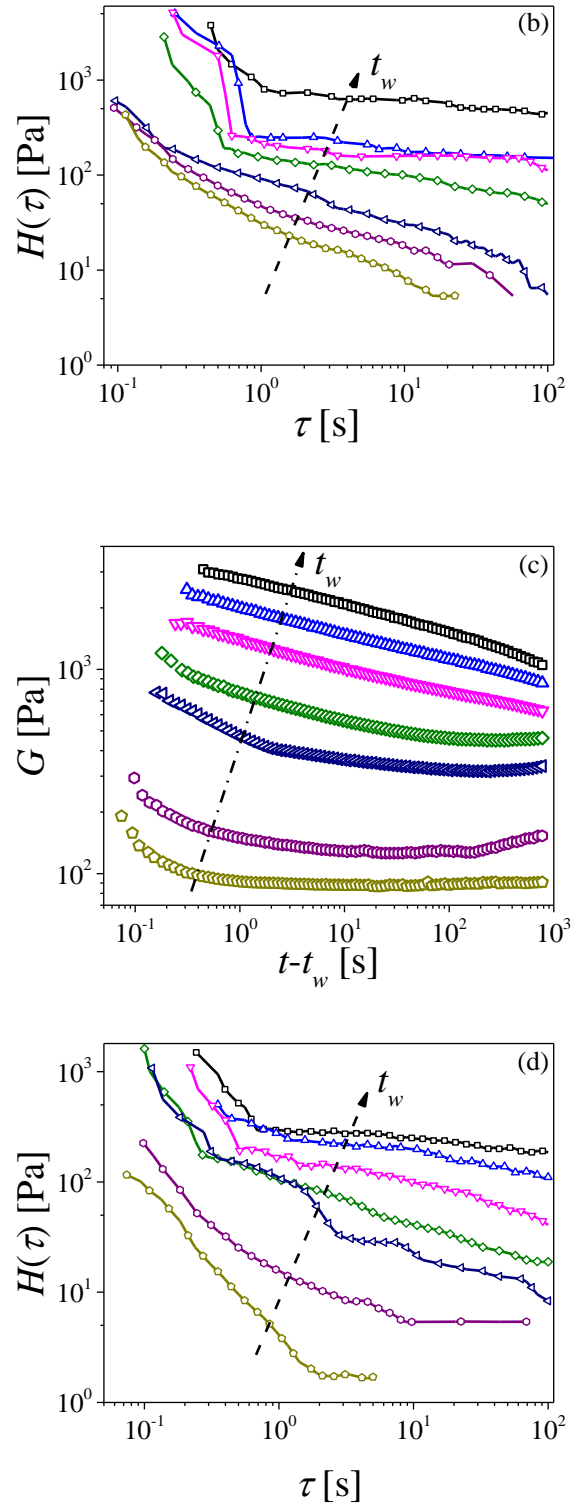

**Fig. A3** Stress relaxation modulus [(a) and (c)] and the corresponding relaxation time spectrum [(b) and (d)] are plotted as a function of time elapsed since application of step strain for a mechanically rejuvenated system with $\gamma_m=200$ % [(a) and (b)] and $\gamma_m=300$ % [(c) and (d)]. The different symbols denote various waiting times $t_w$ after mechanical

rejuvenation: from bottom to top: $t_w$=1800, 3000, 4200, 4800, 6000, 6600, and 8400 s. The arrow is a guide to eye in the increasing direction of $t_w$.

**Acknowledgement**

We acknowledge financial support from the Science and Engineering Research Board, Government of India (Grant number: CRG/2018/003861).

**Author Declarations**

**Conflict of Interest**

The authors have no conflicts of interest to disclose.

**Data Availability**

The experimental data that support the findings of the present study are available from the corresponding author upon reasonable request.